\documentclass[aps,prl,twocolumn,superscriptaddress,floatfix,longbibliography]{revtex4}
\usepackage{amsfonts}
\usepackage{amssymb}
\usepackage{graphicx}
\usepackage{dcolumn}
\usepackage{bm}
\usepackage{amsmath}
\usepackage[colorlinks,linkcolor=magenta,citecolor=blue,urlcolor=blue]{hyperref}
\usepackage{changes}

\begin{document}

\title{Exact mobility edges, $\mathcal{PT}$-symmetry breaking and skin effect in one-dimensional non-Hermitian quasicrystals}
\author{Yanxia Liu}
\affiliation{Beijing National Laboratory for Condensed Matter Physics, Institute of
Physics, Chinese Academy of Sciences, Beijing 100190, China}
\author{Yucheng Wang}
\affiliation{Shenzhen Institute for Quantum Science and Engineering, and Department of
Physics, Southern University of Science and Technology, Shenzhen 518055,
China}
\affiliation{International Center for Quantum Materials, School of Physics, Peking
University, Beijing 100871, China}
\affiliation{Collaborative Innovation Center of Quantum Matter, Beijing 100871, China}
\author{Xiong-Jun Liu}
\affiliation{International Center for Quantum Materials, School of Physics, Peking
University, Beijing 100871, China}
\affiliation{Collaborative Innovation Center of Quantum Matter, Beijing 100871, China}
\author{Qi Zhou}
\email{qizhou@nankai.edu.cn}
\affiliation{Chern Institute of Mathematics and LPMC, Nankai University, Tianjin 300071, China}
\author{Shu Chen}
\email{schen@iphy.ac.cn}
\affiliation{Beijing National Laboratory for Condensed Matter Physics, Institute of
Physics, Chinese Academy of Sciences, Beijing 100190, China}
\affiliation{School of Physical Sciences, University of Chinese Academy of Sciences,
Beijing, 100049, China}
\affiliation{Yangtze River Delta Physics Research Center, Liyang, Jiangsu 213300, China}
\date{\today}
\begin{abstract}
We propose a general analytic method to study the localization transition in one-dimensional quasicrystals with parity-time ($\mathcal{PT}$) symmetry, described by complex quasiperiodic mosaic lattice models. By applying Avila's global theory of quasiperiodic Schr\"odinger operators,  we obtain exact mobility edges and prove that the mobility edge is identical to the boundary of $\mathcal{PT}$-symmetry breaking, which also proves the existence of correspondence between extended (localized) states and  $\mathcal{PT}$-symmetry ($\mathcal{PT}$-symmetry-broken) states. Furthermore, we generalize the models to more general cases with non-reciprocal hopping, which breaks $\mathcal{PT}$ symmetry and generally induces skin effect, and obtain a general and analytical expression of mobility edges. While the localized states are not sensitive to the boundary conditions, the extended states become skin states when the periodic boundary condition is changed to open boundary condition. This indicates that the skin states and localized states can coexist with their boundary determined by the mobility edges.
\end{abstract}

\maketitle




{\it Introduction.-}
The study of localization induced by disorder is a long-standing research
area in condensed matter physics \cite{anderson1958absence}. While localized and extended
states can coexist at different energies in three dimensions with the existence of mobility edges, random disorder generically causes Anderson
localization of the entire spectrum in one and two dimensions \cite{abrahams1979scaling,lee1985disordered,evers2008anderson}. Recently, the interplay of non-Hermiticity and disorder attracted much attention as the non-Hermiticity
brings new perspective for the localization problem by releasing the Hermiticity condition,  e.g., non-Hermitian random matrices
contains  38 different classes according to Bernard-LeClair symmetry classes \cite{BL,HYZhou,CHLiu,Sato}, which generalizes the standard ten classes of Altland-Zirnbauer classification of random Hermitian matrices. In terms of non-Hermitian random-matrix theory, the spectral statistics for non-Hermitian disorder systems has also been unveiled to display some different features from the Hermitian systems \cite{Goldsheid,Markum,Molinari,Chalker}.

The Hatano-Nelson model is a prototype model describing the interplay of the nonreciprocal hopping and random disorder \cite{hatano1996localization,hatano1998non,kolesnikov2000localization,Gong}, which leads to a finite localization-delocalization (LD) transition in one-dimensional (1D) non-Hermitian Anderson model. The effect of complex disorder potentials has also been studied \cite{tzortzakakis2019non,HuangYi}.
Besides the random disorder, the quasiperiodic systems have also attracted intensive studies in recent years \cite{luschen2018,Aubry1980,Kohmoto1983,Thouless1988,roati2008}, including the Aubry-Andr\'{e}
(AA) model \cite{Aubry1980,Kohmoto1983,Thouless1988} and its various extensions \cite{Kohmoto1983,Zhou2013,Cai,DeGottardi,Kohmoto2008}.  By introducing either short-range (long-range) hopping
processes \cite{biddle2011localization,biddle2010predicted,ganeshan2015nearest,li2016quantum,li2017mobility,li2018mobility,DengX}%
or modified quasiperiodic potentials \cite{sarma1988mobility,sarma1990localization,YCWang2020}, the quasiperiodic lattice models  can display energy-dependent mobility edges. While the LD transition in the AA model was analytically predicted in 1980's by utilizing the self-duality property \cite{Aubry1980}, its rigorous mathematical proof was only given recently \cite{Jitomirskaya1999,Avila2008,Avila2017,Avila2015}.
Non-Hermitian quasiperiodic
lattices have also been studied in various references \cite{Yuce,longhi2019metal,jazaeri2001localization,jiang2019interplay,zeng2019topological,longhiPRL,ZengQB}, which mainly focused on the non-Hermitian effect on the AA model. 
Very recently the mobility edges in the non-Hermitian quasicrystal with long-range hopping \cite{Liu2020} or
modified quasiperiodic potentials \cite{Zeng2020,Liutong} have also been studied numerically. 
Although previous works on quasicrystals with $\mathcal{PT}$ symmetry found numerical evidence that the localization transition point coincides with the $\mathcal{PT}$ -symmetry-breaking point \cite{Liu2020,longhiPRL}, the reason behind this observation remains elusive. As nonreciprocal hopping generally induces non-Hermitian skin effect \cite{Yao,Xiong,Kunst,jiang2019interplay,Lee,Alvarez,Yokomizo}, i.e., the exponential accumulation of extended bulk states to edges when the boundary condition is changed from the periodic to open boundary condition (PBC to OBC),  it is not clear whether skin states and localized states can coexist in non-Hermitian quasicrystals and how they are related to the mobility edges?

Aiming to address the above issues, in this letter we first propose a class of $\mathcal{PT}$-symmetrical quasiperiodic mosaic lattices with exact non-Hermitian mobility edges, and rigorously prove the intrinsic relation between the mobility edges and $\mathcal{PT}$ -symmetry breaking by applying Avila's global theory \cite{Avila2015}, one of his Fields Medal work, to the non-Hermitian quasiperiodic system at the first time.
Our method is a general and mathematically rigorous method going beyond the usual dual transformation, which requires some special forms of Hamiltonian \cite{biddle2010predicted,ganeshan2015nearest,Liu2020} to get analytical expression of mobility edges.
Then we study the more general case with nonreciprocal hopping and obtain a concise but unified analytical formula for the mobility edges, which also works as the boundaries separating skin states and localized states.
Our analytical results are crucial to gain exact understanding of the non-Hermitian mobility edges,  $\mathcal{PT}$ -symmetry breaking and interplay of skin effect and  localization in 1D quasicrystals with both complex quasiperiodic potential and non-reciprocal hopping.

{\it Model with $\mathcal{PT}$ symmetry.-}
We consider a 1D quasiperiodic mosaic model with complex quasiperiodic potential
described by%
\begin{equation}
H=\sum_{j}\left( t |j\rangle \left\langle j+1\right\vert +t \left\vert
j+1\right\rangle \langle j|+ V_{j}|j\rangle \langle j|\right) ,
\label{Ham1}
\end{equation}%
with%
\begin{equation}
V_{j}=\left\{
\begin{array}{cc}
2 \lambda \cos (2\pi \omega j+\theta ), & j=\kappa m, \\
0, & otherwise,%
\end{array}%
\right.   \label{lambda}
\end{equation}%
where $\theta =\phi +ih$ describes a
complex phase factor and $\kappa $ is an integer. The quasi-cell has the $%
\kappa $ lattice sites. If the quasi-cell number is taken as $N$, i.e. $m=1$%
, $2$, $\cdots $, $N$, the system size will be $L=\kappa N$. For convenience,
we set $t=1$ as the unit of energy. By taking $|\psi \rangle
=\sum_{j}u_{j}|j\rangle $, the eigen equation is given by $Eu_{j}= u_{j+1}+ u_{j-1}+V_{j}u_{j}$.
Without loss of generality, we take $\omega =\left( \sqrt{5}-1\right) /2$, which can be approached by
$\omega =\lim_{n\rightarrow \infty }%
\frac{F_{n-1}}{F_{n}}$, where $F_{n}$ is the Fibonacci numbers defined rcursively by $F_{n+1}=F_{n}+F_{n-1}$ with $F_{0}=F_{1}=1$.

The model (\ref{Ham1}) has $\mathcal{PT}$ symmetry for $\phi=0$ due to $V_j = V^*
_{-j}$, and its eigenvalues are real if the  $\mathcal{PT}$ symmetry is preserved \cite{Bender}. In the limit of $\lambda=0$, all eigenstates are extended with real eigenvalues. When $\lambda$ increases, localization transition is expected to occur. To study the localization transition, we shall study Lyapunov exponent (LE) of  the model, which can be exactly obtained by applying Avila's global theory of quasiperiodic Schr\"odinger operators \cite{Avila2015}.
The LE can be computed as
\begin{equation}
\gamma \left( E,h\right) = \lim_{n\rightarrow \infty }\frac{1}{n }\int \ln
\left\vert \left\vert T_{n} \left( \phi + i h\right) \right\vert \right\vert
d\phi ,
\end{equation}%
where $\left\vert \left\vert A\right\vert \right\vert $ denotes the norm of
the matrix $A$ and $T_{n}$ is the transfer matrix of a quasi-cell.
Avila's global theory \cite{Avila2015} shows that as a function of $h,$  $\kappa\gamma(E,h)$ is a convex, piecewise linear function, and their slopes are integers.
Thus in  the large-$h$ limit \cite{YCWang2020,SM},  $\kappa\gamma(E,h) =\ln|\lambda a_{\kappa}|+|h|,$ where
\begin{equation}
a_{\kappa } =\frac{1}{\sqrt{E^{2}-4}}\left[ \left( \frac{E+\sqrt{E^{2}-4}}{%
2}\right) ^{\kappa }-\left( \frac{E-\sqrt{E^{2}-4}}{2}\right) ^{\kappa }%
\right] .
  \label{AKa}
\end{equation}

%

Moreover, Avila's global theory can tell us that,  the energy $E$ does not belong to the spectrum of  the Hamilton $H$ with $h=h_0$, if and only if $\gamma(E,h_0)>0$ and $\gamma(E,h_0)$ is an affine functions  in a neighborhood of $h=h_0$.
Consequently, if the energy $E$ lies in the spectrum of the Hamilton $H$ with $h=h_0$, we have
\begin{equation}
\label{Lyapunovh}
\kappa \gamma(E,h_0)= \max\{\ln| \lambda a_{\kappa}(E)|+|h_0|, 0\}.
\end{equation}
Specially, when $h_0=0$, $ \kappa \gamma(E,0)= \max\{\ln| \lambda a_{\kappa}(E)|, 0\}.$
Note that $\gamma \left( E,h_0\right) =0$ corresponds to the extended state,  $\gamma \left( E,h_0\right) >0$ indicates the localized state, which gives rise to the mobility edge determined by
\begin{equation}
e^{\left\vert h \right\vert} \left\vert a_{\kappa }(E_c) \right\vert = \frac{1}{\left\vert \lambda \right\vert}\text{,}  \label{ME1}
\end{equation}%
where $a_{\kappa }$ is given by Eq.(\ref{AKa}) and $h=h_0$. \begin{figure}[tbp]
\includegraphics[width=0.45\textwidth]{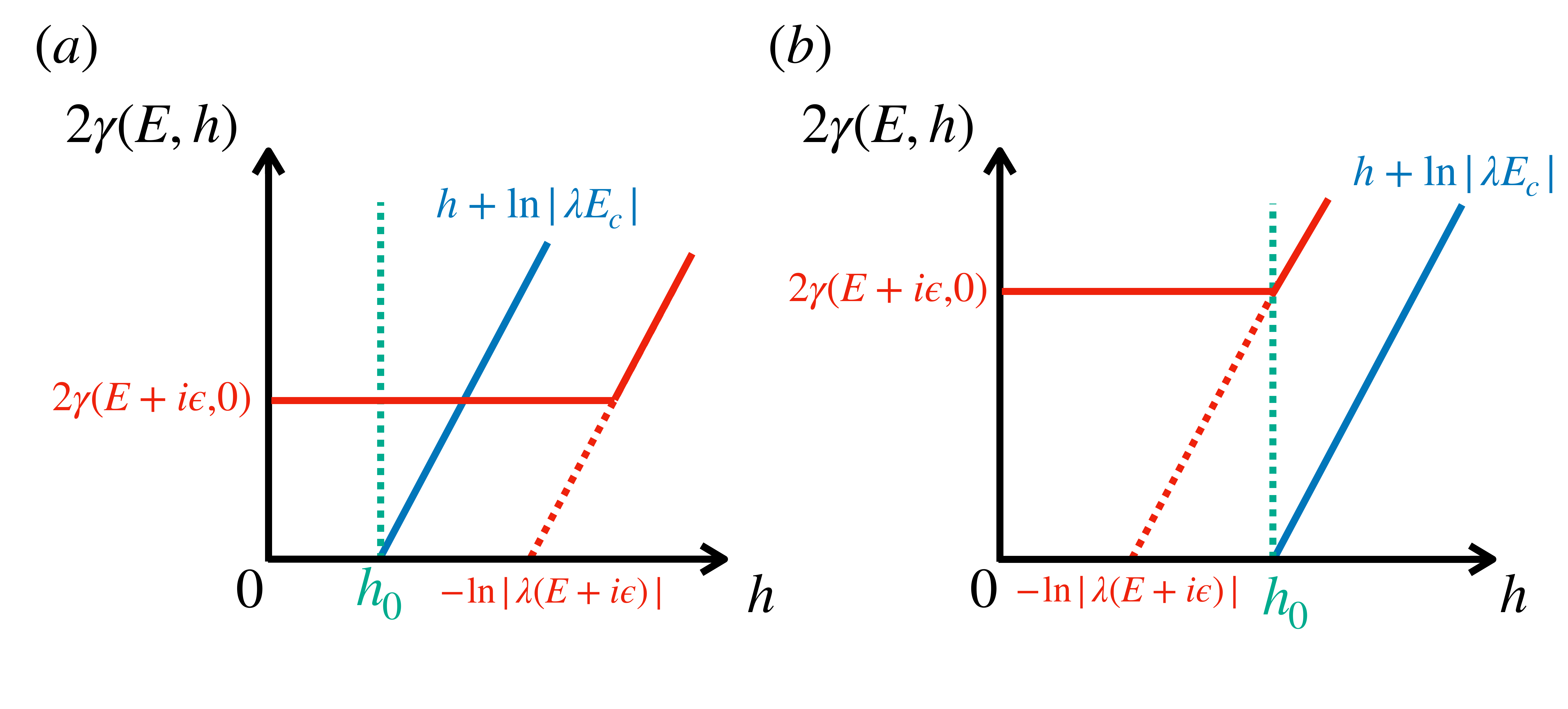}
\caption{ The blue line shows the Lyapunov exponent $\gamma(E_c,h)$ for critical energy $E_c$, where $E_c$ is the
mobility edge for the system with $h=h_0=-\ln |\lambda E_c|$.  The red line shows the Lyapunov exponent $\gamma(E+i\epsilon,h)$.  (a) No complex energy in the regime  $|E+i\epsilon|<E_c$ belong to the spectrum of the system with $h=h_0$. (b)  In the regime  $|E+i\epsilon|>E_c$, the complex energy might belong to  the spectrum of the system with $h=h_0$,  when $\gamma(E+i\epsilon,h_0)$ is an extreme point of $\gamma(E+i\epsilon,h)$. }
\label{fig1}
\end{figure}

If  $\kappa=2$, then $a_{2}(E)=E$, and
\begin{equation}
E_c=  1/(\lambda e^{|h|}). \label{MEE}
\end{equation}
By the above discussion especially formula \eqref{Lyapunovh}, one can conclude that the eigenstates with energies $|E
+i\epsilon|<E_c$ and $|E+i\epsilon|>E_c$
correspond to the extended states and localized states, respectively.
Now we analysis the distribution of the real energies and complex energies of the system with $h=h_0$.
To that end, we should start with the case $h=0$.
If $h=0$, then the Hamiltonian is Hermitian, and the spectrum is real. The complex energy $E+i\epsilon$ does not belong to the spectrum of
the system with $h=0$, so $\gamma(E+i\epsilon,0)>0$, where $E$ and $\epsilon$ are real. If $|E+i\epsilon|<E_c$, it's easy
to see that $ \gamma(E+i\epsilon,h)$ is an affine functions in a neighborhood of $h=h_0$, so in this regime, complex energy $E+i\epsilon$
does not belong to the spectrum of the system with $h=h_0$ either, as shown in Fig.\ref{fig1}(a). The extended states only happens for real energies, which possess $\mathcal{PT}$ symmetry. On the other hand,  $E+i\epsilon$  belongs to the spectrum of the system with $h=h_0$, if and only if $$ |h_0|+\ln |\lambda (E+i\epsilon) |=2\gamma(E+i\epsilon,0),$$ as shown in Fig.\ref{fig1}(b), where $|E+i\epsilon|>E_c$. The complex energies only correspond to the localized states, which have no $\mathcal{PT}$-symmetry.  We stress that our analytic reasoning applies to general $\kappa$.

\begin{figure}[tbp]
\includegraphics[width=0.45\textwidth]{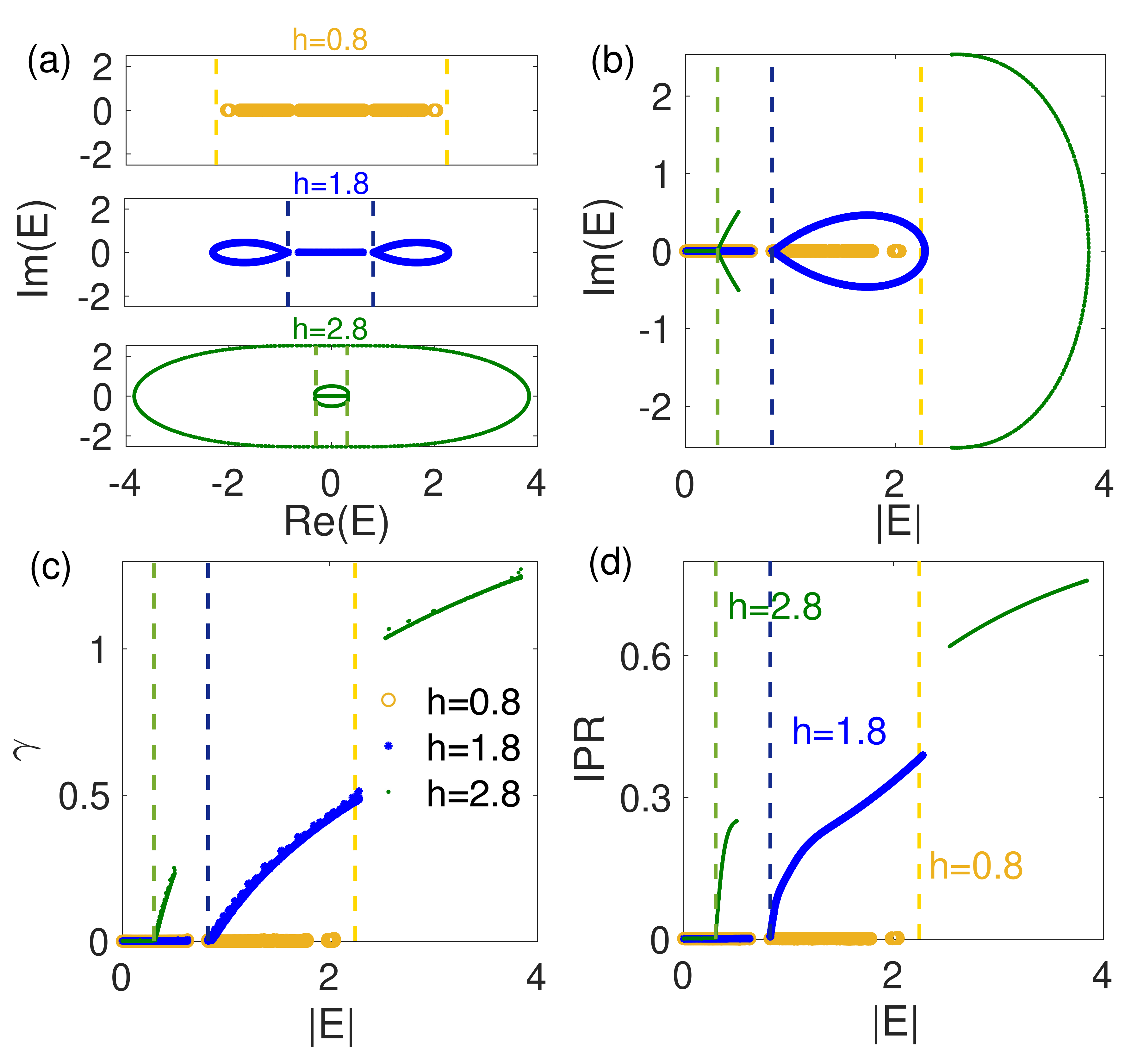}
\caption{Eigenvalues in the space spanned by (a) Im$(E)$ and Re$(E)$ and
(b) Im$(E)$ and $|E|$, (c) the Lyapunov exponent, and
(d) IPR versus $|E|$ for the system with $\lambda=0.2$, $h=0.8$, $1.8$ and $2.8 $ under PBC.
The dashed lines represent the mobility edges obtained by Eq.(\ref{MEE}). The
quasi-cell number is set to be $N=233$.}
\label{fig2}
\end{figure}

To get an intuitive understanding of the above analytical results, we shall demonstrate numerical results of the energy spectrum, LE and inverse participation ratio (IPR) for the system with $\kappa =2$.
For a finite-site lattice, the LE can be obtained by numerically calculating
$
\gamma \left( E\right) =\ln \left( \max \left( \theta _{i}^{+},\theta
_{i}^{-}\right) \right)
$,
where $\theta _{i}^{\pm }\in \mathbb{R}$ are the eigenvalues of the matrix $\mathbf{\Theta =}\left( T_{\kappa N}^{\dag }\left(E, \phi ,h\right)
T_{\kappa N}\left(E, \phi ,h\right) \right) ^{1/(2L)}$. There are two
eigenvalues and the LE is taken to be the maximum one.
The IPR of an eigenstate is defined as  $\text{IPR}^{(i)}= (\sum_{n}\left\vert u_{n}^{i}\right\vert ^{4})/ \left(
\sum_{n}\left\vert u_{n}^{i}\right\vert ^{2}\right) ^{2}$,
where the superscript $i$ denotes the $i$th eigenstate, and $n$
labels the lattice site.
While IPR$\simeq 1$ for a full localized eigenstate, IPR$\simeq 1/L$ for an extended eigenstate and tends to zero as $L \rightarrow \infty$.

Figs.\ref{fig2}(a)-(d) show numerical results of systems with fixed $\lambda =0.2$ and different $h$. Here the eigenvalues are plotted in the complex space spanned by Im$(E)$ and Re$(E)$  in Fig.\ref{fig2}(a), and also mapped to the space spanned by Im$(E)$ and $|E|$ in Fig.\ref{fig2}(b), in order to compare with Fig.\ref{fig2}(c) and  Fig.\ref{fig2}(d), which display LE and IPR versus $|E|$, respectively.
For $h=0.8$, all eigenvalues are real with $|E|<|E_c|$, which indicates the corresponding eigenstates being extended states. In this case, no $\mathcal{PT}$-symmetry breaking happens as all eigenstates with real eigenvalues fulfill the $\mathcal{PT}$-symmetry. When $h$ exceeds a critical value $h_c=- \ln (\lambda E_{max}) \approx 0.9$, where $E_{max}$ is the maximum eigenvalue of the corresponding Hermitian Hamiltonian,  $\mathcal{PT}$-symmetry breaking happens. While eigenvalues fulfilling $|E|<|E_c|$ are still real,  they become complex when $|E| > |E_c|$ as displayed in Figs.\ref{fig2}(a) and \ref{fig2}(b) for cases of $h=1.8$ and $2.8$. Both the LE and IPR have a sudden increase when $|E| > |E_c|$,  which confirms mobility edges are consistent with the analytical results.  Although increasing $h$ shall shrink $E_c$,
the real extended states and complex localized states always coexist even $h \rightarrow \infty$. Such a novel phenomenon was never predicted in literature.
Our results demonstrate that 
the transition from extended to localized states and  $\mathcal{PT}$-symmetry breaking transition have the same boundary.
So 
the LD transition can be also read out from the change of spectrum structure.

{\it Model with non-reciprocal hopping.-}
Now we consider a more general case with Hamiltonian given by
\begin{equation}
\tilde{H}=\sum_{j}\left( t_{L}|j\rangle \left\langle j+1\right\vert +t_{R}\left\vert
j+1\right\rangle \langle j|+ V_{j}|j\rangle \langle j|\right) ,
\label{Ham-nonrecip}
\end{equation}%
where $t_{L}=te^{-g}$ and $t_{R}=te^{g}$ are the left-hopping and
right-hopping amplitude, respectively, and $V_{j}$ is given by Eq.(\ref{lambda}).  The nonreciprocal hopping breaks the $\mathcal{PT}$ symmetry of Hamiltonian, and also  generally induces skin effect under OBC. The Hamiltonian $\tilde{H}(g)$ under OBC can be transformed to $H$ via a similar transformation
$H =S  \tilde{H}(g) S^{-1}$,
where $S=$diag$\left( e^{-g},e^{-2g},\cdots ,e^{-Ng}\right) $ is a
similarity matrix with exponentially decaying diagonal entries and $H  = \tilde{H}(g=0)$ is just the Hamiltonian (\ref{Ham1}) under OBC. The eigenvectors of $\tilde{H}$ and $H$ satisfy $\left\vert \tilde{\psi} \right\rangle =S^{-1}\left\vert \psi
\right\rangle$.
An extended states $\left\vert \psi \right\rangle $ under the transformation $S^{-1}$ becomes skin-mode states,
which are exponentially accumulated to one boundary. A localized
state of $H$ generally takes the form $\left\vert u_{j}\right\vert \propto e^{-\gamma
\left\vert j-j_{0}\right\vert }$, where $j_{0}$ is the index of the
localization center, and $\gamma $ is the Lyapunov exponent in (\ref{Lyapunovh}).
Then the corresponding wavefunction of $\tilde{H}(g)$ takes the form of
\begin{equation*}
\left\vert u_{j}\right\vert \propto \left\{
\begin{array}{cc}
e^{-\left( \gamma -g\right) \left\vert j-j_{0}\right\vert } & j>j_{0} \\
e^{-\left( \gamma +g\right) \left\vert j-j_{0}\right\vert } & j<j_{0}%
\end{array}%
\right. ,  
\end{equation*}%
which manifest different decaying behaviors on two sides of the localization
center. When $%
 \left\vert g\right\vert \geq \gamma $, delocalization occurs on one side and
then skin modes emerge to the boundary on the same side. The transition point of localized states and skin states is given by
\begin{equation}
\frac{\left\vert h\right\vert +\ln \left\vert \lambda a_{\kappa }\right\vert
}{\kappa }=\left\vert g\right\vert .\label{LDP}
\end{equation}%
Since the localized state is not sensitive to the boundary condition, we can conclude the LD transition in the periodic boundary system is also given by Eq.(\ref{LDP}), which gives rise to the mobility edge
\begin{equation}
\left\vert \lambda a_{\kappa } (E_c) \right\vert
=  e^{\kappa \left\vert g\right\vert - \left\vert h\right\vert } . \label{mobt2}
\end{equation}%
For $\kappa =2$,  the mobility edge is given by
\begin{equation}
\left\vert
\lambda E_c \right\vert = e^{ 2\left\vert g\right\vert -\left\vert h\right\vert}.  \label{mob2}
\end{equation}%
Since the eigenvalue of $\tilde{H}$ is generally complex, the mobility edge is only associated with the absolute value of the eigenvalue. While increasing $h$ suppresses $|E_c|$, $|g|$ tends to enlarge the region of extended states. Particularly, when $|h|=2|g|$, $|E_c| =1/|\lambda|$.

\begin{figure}[tbp]
\includegraphics[width=0.5\textwidth]{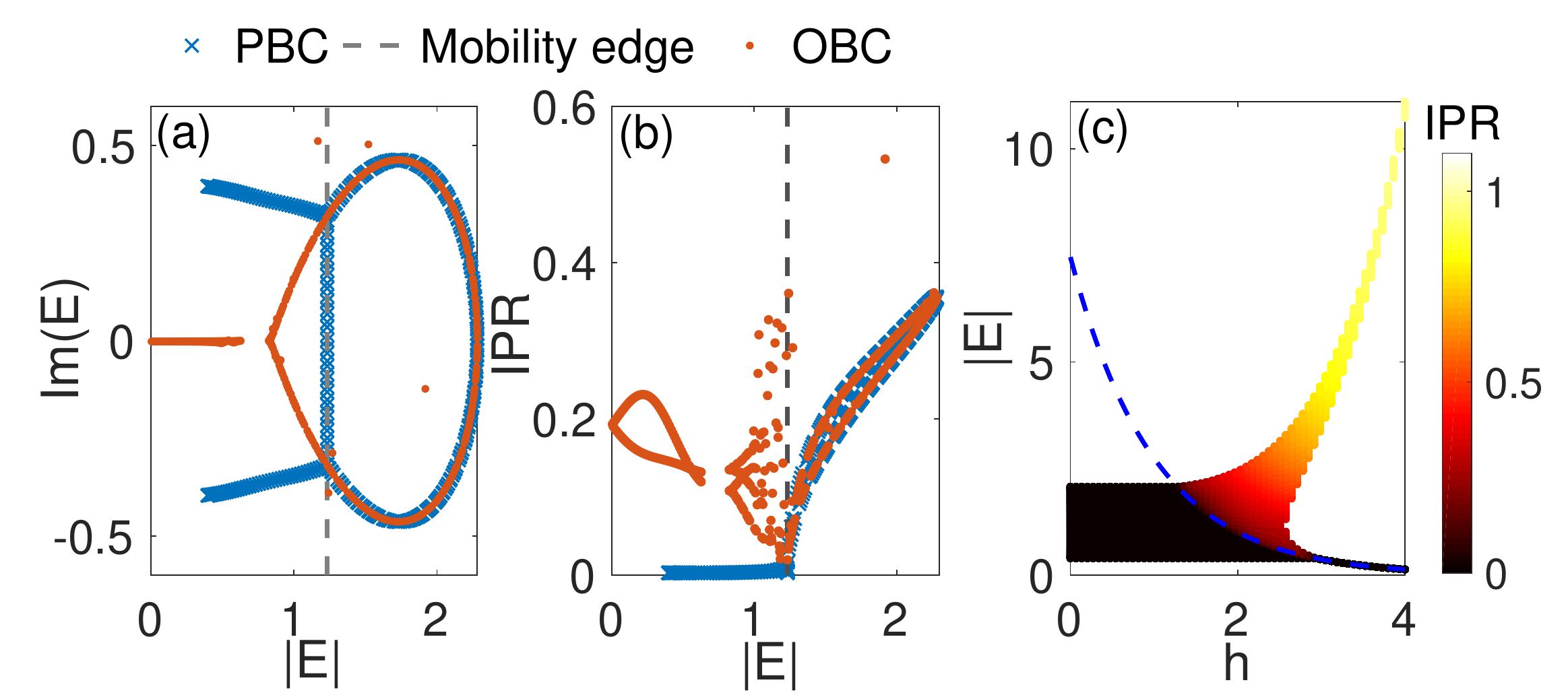}
\caption{(a) Eigenvalues  in the space spanned by Im$E$ and $|E|$ and (b) IPR versus
$|E|$ for the system with $\lambda=0.2$, $g=0.2$, $h=1.8$ and $N=233$ under PBC (red  dots) and OBC (blue crosses). (c) (b) IPR of different eigenstates as a function
of the corresponding absolute value of
eigenenergies and $h$ with  $g=0.2$ and $\lambda=0.2$. Dashed lines represent
mobility edges.}
\label{fig3}
\end{figure}

The similar transformation suggests that the eigenvalue of $\tilde{H}$ under OBC is identical to $H$, i.e., the open boundary eigenvalue is irrelevant with $g$. On the other hand, the spectrum under PBC depends on $g$, which is clearly manifested by the periodic spectrum of  $E=2t \cos (k+ig)$ in the limit of $\lambda=0$. Since the similar transition only holds true under the OBC, the spectrum of $\tilde{H}$ under the OBC and PBC are generally different \cite{SM}. In Fig.\ref{fig3}(a), we display the spectrum of the system under both PBC and OBC. While eigenvalues with $|E|>|E_c|$ are shown to be almost the same under both PBC and OBC, the parts of spectra with $|E|<|E_c|$ are obviously different under different boundaries, which is a character of the existence of skin effect \cite{KZhang,Okuma}.
In Fig.\ref{fig3}(b), we plot the IPR versus $|E|$ for both systems under PBC and OBC. The IPRs have a sudden increase when $|E|>|E_c|$ and display almost the same distributions in this localized region under different boundary conditions. Due to the existence of skin states in the region of $E<|E_c|$, the IPRs of open boundary system take finite values and are obviously different from the periodic system.  In Fig.\ref{fig3}(c), we display IPRs of the corresponding eigenstates
as a function of $h$ for the periodic system with $\lambda=0.2$ and $g=0.2$. The dashed lines in Fig.\ref{fig3}
represent the mobility edges determined by Eq.(\ref{mob2}), which separate the extended
and localized states with the values of IPR below which approaching zero and
above being finite. The numerical results from IPR agree well with the
analytical relation given by Eq. (\ref{mob2}).

\begin{figure}[tbp]
\includegraphics[width=0.45\textwidth]{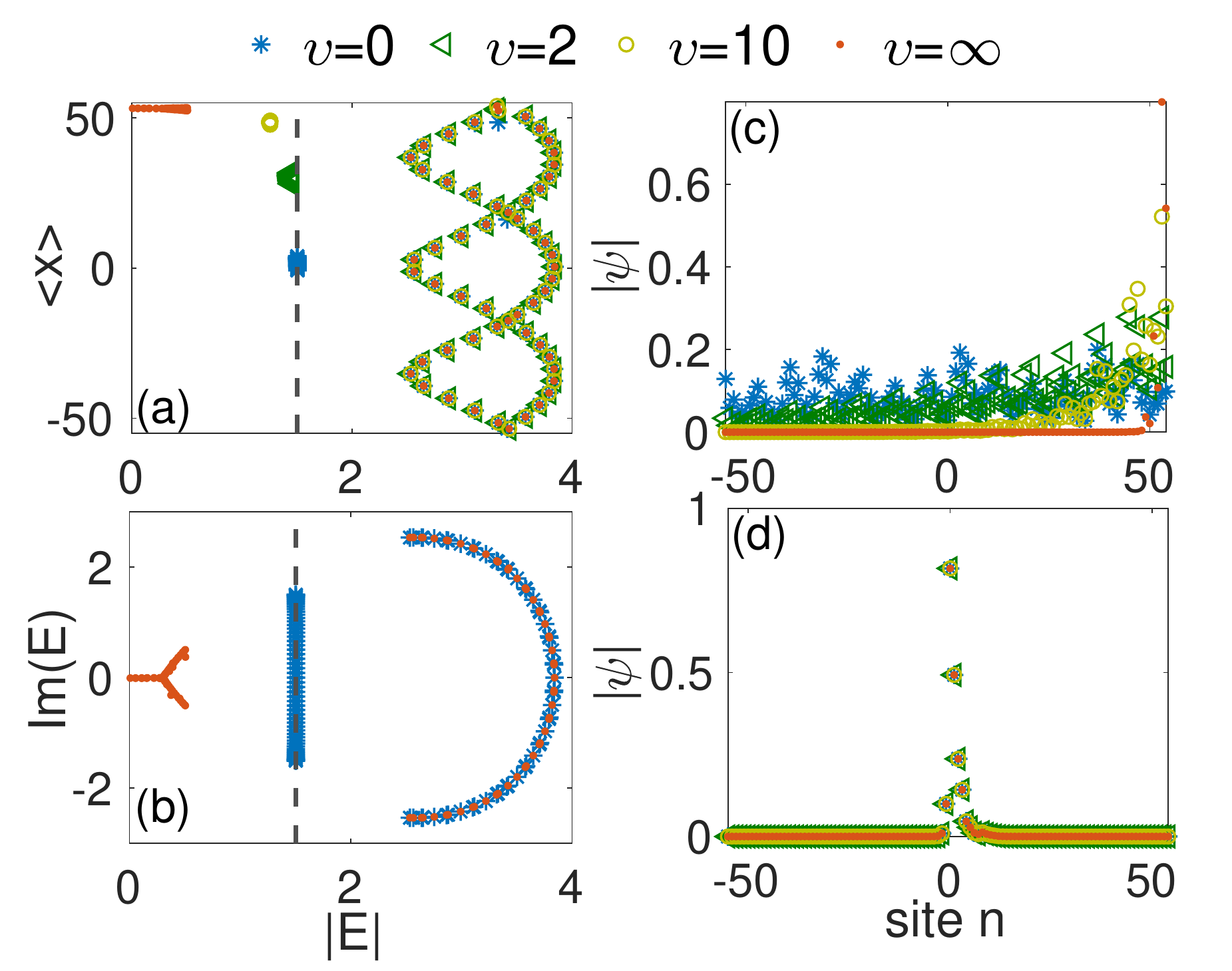}
\caption{(a) The mean position of wavefunctions $\langle x \rangle$ for the system with $\lambda=0.2$,
$g=0.8$, $h=2.8$, $N=55$ and different $\upsilon$. (b)
Eigenvalues  in the space spanned by Im$E$ and $|E|$ for the same system under OBC and PBC.
The dashed lines represent the mobility edges.
The distribution of wavefunction $|\protect\psi|$ corresponding to the minimum (c)
and maximum (d) $\left\vert E\right\vert $ with $\upsilon=0$, $2$
, $10$, $\infty$.}
\label{fig4}
\end{figure}

The insensitivity of the localized states to the boundaries has suggested that the onset of localization transition should be irrelevant to the boundary conditions. To see the effect of boundary clearly, we consider that the hopping term between the $L$-th and first site is replaced by
$h_{L1} = \eta \left( t_{L}|L\rangle \left\langle 1\right\vert +t_{R}\left\vert1\right\rangle \langle L|\right)$,
with  the introduction of a boundary anisotropic parameter $\eta \in [0,1]$ \cite{Lee}. For convenience, we take $\eta=e^{-\upsilon}$ with $\upsilon=0$ $\left( \infty \right) $ corresponding to
the PBC (OBC).  In Fig.{\ref{fig4}}(a), we display
the mean position $\left\langle x\right\rangle =\left\langle
\psi \right\vert \hat{x} |\psi \rangle $ of eigenstates versus $|E|$ under different boundary conditions, where $\hat{x} =\sum_n n |n\rangle \langle n|$ is the position operator. Fig.{\ref{fig4}}(b) shows the corresponding spectrum in the parameter space spanned by $Im(E)$ and $|E|$.
While $\left\langle x\right\rangle$ near the center of the
lattices indicates that the wavefunction distributes over the whole lattice (the
extended state), its accumulation on one of the boundary
corresponds to skin states. For
the localized states, $\left\langle x\right\rangle $ can take arbitrary values
within position of lattices. It is shown that the extended states are sensitive to the boundary condition and become skin states under the OBC, whereas the localized
states almost have no change with the change of boundary anisotropic parameter. This is also witnessed by wavefunction distributions shown in Figs.{\ref{fig4}}(c) and {\ref{fig4}}(d).

{\it Summary and discussion.-}
In summary,  we proposed a general analytic method to study the LD transition and $\mathcal{PT}$-symmetry breaking  for  non-Hermitian quasiperiodic models. Specially, we studied 1D non-Hermitian quasiperiodic mosaic models with both complex quasiperiodic potential and non-reciprocal hopping and obtained analytically
the exact mobility edges uniformly described by Eq.(\ref{mobt2}), which is the central result of the present work. For the case with $\mathcal{PT}$ symmetry, we proved that the mobility edge is identical to the boundary of $\mathcal{PT}$-symmetry breaking. In the presence of non-reciprocal hopping, while the localized states are not sensitive to the boundary conditions, extended states
are driven to skin states when the PBC is changed to OBC, and skin states can coexist with localized states with their boundaries given by the mobility edges.

While the mobility edges only exist for cases of $\kappa \geq 2$, we note Eq.(\ref{mobt2}) still holds true for  $\kappa =1$, where $a_{1} =1 $ and the model reduces to the non-Hermitian AA models.
From Eq.(\ref{mobt2}), we get the localization transition occurring at
$
|\lambda| =e^{-|h|+|g|}
$.
It is clear that the transition point is irrelevant to the eigenvalue $E$, indicating that no mobility
edges exist. All eigenstates are extended (localized) when $|\lambda| < e^{-|h|+|g|}$ ($|\lambda|> e^{-|h|+|g|}$), 
which recovers the result of Ref.\cite{longhiPRL} for $h\neq 0$ and $g=0$ and the result of Ref.\cite{jiang2019interplay} for $g\neq 0$ and $h=0$.
For $h=g=0$, our model reduces to its Hermitian limit \cite{YCWang2020,Aubry1980}.
An interesting limit case of our model is obtained in the double limit $h \rightarrow \infty$, $\lambda \rightarrow 0$ with $\lambda e^h \rightarrow V$ finite, corresponding to the quasiperiodic potential given by $V_j = V \exp(-i 2 \pi \omega j)$ for $j=km$ and $0$ otherwise in Eq.(\ref{Ham-nonrecip}). In this limit, the mobility edges are given by
$ \left\vert  a_{\kappa } (E_c) \right \vert = e^{\kappa \left\vert g\right\vert}  /|V|$.
The diversity and solvability of our models provide a new zoo for analytically exploring the richness of non-Hermitian localization phenomena.

\begin{acknowledgments}
This work is supported by NSFC under Grants Nos. 11974413, the National
Key Research and Development Program of China (2016YFA0300600 and 2016YFA0302104) and the Strategic Priority Research Program of CAS (XDB33000000). Q. Zhou was partially supported by support by NSFC grant (11671192, 11771077), The Science Fund for Distinguished Young Scholars of Tianjin (No. 19JCJQJC61300) and Nankai Zhide Foundation.
Y. Wang and X.-J. Liu are supported by National Nature Science Foundation of China (11825401, 11761161003, and 11921005), the National Key R\&D Program of China (2016YFA0301604), Guangdong Innovative and Entrepreneurial Research Team Program (No.2016ZT06D348), the Science, Technology and Innovation Commission of Shenzhen Municipality (KYTDPT20181011104202253), and the Strategic Priority Research Program of Chinese Academy of Science (Grant No. XDB28000000).
\end{acknowledgments}


\begin{thebibliography}{99}

\bibitem{anderson1958absence} P. W. Anderson, Absence of diffusion in certain random lattices, Phys. Rev. \textbf{109}, 1492(1958).

\bibitem{abrahams1979scaling} E. Abrahams, P. W. Anderson, D. C. Licciardello, and T. V. Ramakrishnan,
Scaling theory of localization: Absence of quantum diffusion in two dimensions, Phys. Rev. Lett. \textbf{42}, 673 (1979).

\bibitem{lee1985disordered} P.  A. Lee and T. V. Ramakrishnan, Disordered
electronic systems, Rev. Mod. Phys. \textbf{57}, 287(1985).


\bibitem{evers2008anderson} F. Evers and A. D. Mirlin, Anderson
transitions, Rev. Mod. Phys. \textbf{80}, 1355 (2008).

\bibitem{BL} D. Bernard and A. LeClair, A classification of non-Hermitian random matrices, arXiv:0110649.

\bibitem{Sato} K. Kawabata, K. Shiozaki, M. Ueda, and M. Sato, Symmetry and
topology in non-Hermitian physics, Phys. Rev. X \textbf{9}, 041015 (2019).

\bibitem{HYZhou} H. Zhou and J. Y. Lee, Periodic table for topological bands with non-Hermitian symmetries, Phys. Rev. B \textbf{99}, 235112
(2019).

\bibitem{CHLiu} C.-H. Liu, and S. Chen, Topological classification of defects
in non-Hermitian systems, Phys. Rev. B \textbf{100}, 144106 (2019).


\bibitem{Goldsheid}I. Y. Goldsheid and B. A. Khoruzhenko,
Distribution of Eigenvalues in Non-Hermitian Anderson Models, Phys. Rev. Lett. {\bf 80} 2897 (1998).

\bibitem{Molinari} L. G. Molinari, Non-Hermitian spectra and Anderson localization, J. Phys. A: Math. Theor. {\bf 42} 265204 (2009).


\bibitem{Markum} H. Markum, R. Pullirsch, and T. Wettig, Non-Hermitian Random Matrix Theory and Lattice QCD with Chemical Potential,
Phys. Rev. Lett. {\bf 83}, 484 (1999).


\bibitem{Chalker} J. T. Chalker and B. Mehlig, Eigenvector Statistics in Non-Hermitian Random Matrix Ensembles,
Phys. Rev. Lett. {\bf 81}, 3367 (1998).



\bibitem{hatano1996localization} N. Hatano and D. R. Nelson, Localization
transitions in non-hermitian quantum mechanics, Phys. Rev. Lett. \textbf{77}%
, 570 (1996).

\bibitem{hatano1998non} N. Hatano and D. R. Nelson, Non-hermitian
delocalization and eigenfunctions, Phys. Rev. B \textbf{58}, 8384 (1998).


\bibitem{kolesnikov2000localization} A. V. Kolesnikov and K. B. Efetov,
Localization- delocalization transition in non-hermitian disordered systems,
Phys. Rev. Lett. \textbf{84}, 5600 (2000).


\bibitem{Gong} Z. Gong, Y. Ashida, K. Kawabata, K. Takasan, S. Higashikawa,
and M. Ueda, Topological Phases of Non-Hermitian Systems, Phys. Rev. X
\textbf{8}, 031079 (2018).

\bibitem{tzortzakakis2019non} A. F. Tzortzakakis, K. G. Makris, and E. N. Economou, Non-hermitian disorder in two-dimensional optical lattices, Phys. Rev. B \textbf{101}, 014202 (2020).

\bibitem{HuangYi} Y. Huang and B. I. Shklovskii, Anderson transition in
three-dimensional systems with non-Hermitian disorder, Phys. Rev. B \textbf{%
101}, 014204 (2020).

\bibitem{roati2008} G. Roati, C. DErrico, L. Fallani, M. Fattori, C.
Fort, M. Zaccanti, G. Modugno, M. Modugno, and M. Inguscio, Anderson
localization of a non-interacting bose Ceinstein condensate, Nature
(London) \textbf{453}, 895 (2008).


\bibitem{luschen2018} H. P. L\"uschen, S. Scherg, T. Kohlert, M.
Schreiber, P. Bordia, X. Li, S. Das Sarma, and I. Bloch, Single-particle
mobility edge in a one-dimensional quasiperiodic optical lattice, Phys. Rev.
Lett. \textbf{120}, 160404 (2018).


\bibitem{Aubry1980} S. Aubry and G. Andr\'{e}, Analyticity breaking and
Anderson localization in incommensurate lattices, Ann. Israel Phys. Soc.
\textbf{3}, 133 (1980).

\bibitem{Kohmoto1983} M. Kohmoto, Metal-insulator transition and scaling for
incommensurate systems, Phys. Rev. Lett. \textbf{26}, 1198 (1983).

\bibitem{Thouless1988} D. J. Thouless, Localization by a
potential with slowly varying period, Phys. Rev. Lett. \textbf{61},
2141(1988).


\bibitem{Zhou2013} L. Zhou, H. Pu, and W. Zhang, Anderson localization of
cold atomic gases with effective spin-orbit interaction in a quasiperiodic
optical lattice, Phys. Rev. A \textbf{87}, 023625 (2013).

\bibitem{Kohmoto2008} M. Kohmoto and D. Tobe, Localization problem in a
quasiperiodic system with spin-orbit interaction, Phys. Rev. B \textbf{77},
134204 (2008).

\bibitem{Cai} X. Cai, L.-J. Lang, S. Chen, and Y. Wang, Topological superconductor
to Anderson localization transition in one-Dimensional incommensurate lattices,
Phys. Rev. Lett. {\bf 110}, 176403 (2013).

\bibitem{DeGottardi} W. DeGottardi, D. Sen, and S. Vishveshwara, Majorana fermions
in superconducting 1D systems having periodic, quasiperiodic, and disordered Potentials, Phys. Rev. Lett.
{\bf 110}, 146404 (2013).


\bibitem{biddle2011localization} J. Biddle, D. J. Priour, B. Wang, and S.
Das Sarma, Localization in one-dimensional lattices with
non-nearest-neighbor hopping: Generalized Anderson and Aubry- Andr\'e
models, Phys. Rev. B \textbf{83}, 075105 (2011).

\bibitem{biddle2010predicted} J. Biddle and S. Das Sarma, Predicted mobility
edges in one-dimensional incommensurate optical lattices: An exactly
solvable model of Anderson localization, Phys. Rev. Lett. \textbf{104},
070601 (2010).

\bibitem{ganeshan2015nearest} S. Ganeshan, J. H. Pixley, and S. Das Sarma,
Nearest neighbor tight binding models with an exact mobility edge in one
dimension, Phys. Rev. Lett. \textbf{114}, 146601 (2015).

\bibitem{li2016quantum} X. P. Li, J. H. Pixley, D. L. Deng, S. Ganeshan, and
S. Das Sarma, Quantum nonergodicity and fermion localization in a system
with a single-particle mobility edge, Phys. Rev. B \textbf{93}, 184204
(2016).

\bibitem{li2017mobility} X. Li, X. P. Li, and S. Das Sarma, Mobility edges
in one-dimensional bichromatic incommensurate potentials, Phys. Rev. B
\textbf{96}, 085119 (2017).

\bibitem{li2018mobility} X. Li and S. Das Sarma, Mobility edge and
interme-diate phase in one-dimensional incommensurate lattice potentials,
Phys. Rev. B \textbf{101}, 064203 (2020).

\bibitem{DengX} X. Deng, S. Ray, S. Sinha, G. V. Shlyapnikov, and L. Santos,
One-Dimensional Quasicrystals with Power-Law Hopping, Phys. Rev. Lett.
\textbf{123}, 025301 (2019).

\bibitem{sarma1988mobility} S. Das Sarma, S. He, and X. C. Xie, Mobility
edge in a model one-dimensional potential, Phys. Rev. Lett. \textbf{61},
2144(1988).

\bibitem{sarma1990localization} S. Das Sarma, S. He, and X. C. Xie,
Localization, mobility edges, and metal-insulator transition in a class of
one-dimensional slowly varying deterministic potentials, Phys. Rev. B
\textbf{41}, 5544 (1990).

\bibitem{YCWang2020} Y. Wang, X. Xia, L. Zhang, H. Yao, S. Chen, J. You, Q.
Zhou, and X. Liu, One dimensional quasiperiodic mosaic lattice with exact
mobility edges, arXiv:2004.11155.


\bibitem{Jitomirskaya1999} S. Y. Jitomirskaya, Metal-insulator transition
for the almost mathieu operator, Ann. Math. \textbf{3}, 150 (1999).

\bibitem{Avila2008} A. Avila, The absolutely continuous spectrum of the
almost Mathieu operator, arXiv:0810.2965.

\bibitem{Avila2017} A. Avila, J. You , Q. Zhou, Sharp phase transitions for
the almost Mathieu operator, Duke. Math. J. \textbf{14}, 166 (2017)
\bibitem{Avila2015} A. Avila, Global theory of one-frequency Schr\"{o}inger
operators, Acta. Math. \textbf{1}, 215, (2015).

\bibitem{jazaeri2001localization} A. Jazaeri and I. I. Satija, Localization
transition in incommensurate non-hermitian systems, Phys. Rev. E \textbf{63}%
, 036222 (2001).

\bibitem{Yuce} C. Yuce. $\mathcal{PT}$ symmetric Aubry-Andre model, Phys. Lett. A \textbf{378}, 2024 (2014).

\bibitem{ZengQB} Q.-B. Zeng, S. Chen, and R. Lu, Anderson localization in
the non-Hermitian Aubry-Andre-Harper model with physical gain and loss,
Phys. Rev. A \textbf{95}, 062118 (2017).

\bibitem{longhi2019metal} S. Longhi, Metal-insulator phase transition in a
non-hermitian aubry-andr\'e-harper model, Phys. Rev. B \textbf{100}, 125157
(2019).

\bibitem{longhiPRL} S. Longhi, Topological phase transition in
non-hermitian quasicrystals, Phys. Rev. Lett. \textbf{122}, 237601 (2019).

\bibitem{jiang2019interplay} H. Jiang, L. J. Lang, C. Yang., S. L. Zhu, and
S. Chen, Interplay of non-hermitian skin effects and anderson localization
in nonreciprocal quasiperiodic lattices, Phys. Rev. B \textbf{100}, 054301
(2019).

\bibitem{zeng2019topological} Q. B. Zeng, Y. B. Yang, and Y. Xu, Topological
phases in non-hermitian aubry-andr\'e-harper models, Phys. Rev. B \textbf{101%
}, 020201 (2020).

\bibitem{Liu2020} Y. Liu, X.-P. Jiang, J. Cao, and S. Chen, Non-Hermitian
mobility edges in one-dimensional quasicrystals with parity-time symmetry,
Phys. Rev. B \textbf{101}, 174205 (2020).

\bibitem{Zeng2020} Q.-B. Zeng and Y. Xu, Winding numbers and generalized
mobility edges in non-Hermitian systems, Phys. Rev. Research
{\bf 2}, 033052 (2020).

\bibitem{Liutong} T. Liu, H. Guo, Y. Pu, and S. Longhi, Generalized Aubry-Andre self-duality and mobility edges in non-Hermitian quasiperiodic lattices
Phys. Rev. B {\bf 102}, 024205 (2020).

\bibitem{Yao} S. Yao and Z. Wang, Edge states and topological invariants of
non-Hermitian systems, Phys. Rev. Lett. \textbf{121}, 086803 (2018).

\bibitem{Xiong} Y. Xiong, Why does bulk boundary correspondence fail in some
non-Hermitian topological models, J. Phys. Commun. \textbf{2}, 035043 (2018).

\bibitem{Alvarez} V. M. Martinez Alvarez, J. E. Barrios Vargas, and L. E. F. Foa
Torres, Non-Hermitian robust edge states in one dimension:
Anomalous localization and eigenspace condensation at exceptional
points, Phys. Rev. B {\bf 97}, 121401(R) (2018).

\bibitem{Kunst} F. K. Kunst, E. Edvardsson, J. C. Budich, and E. J.
Bergholtz, Biorthogonal, bulk-boundary correspondence in non-Hermitian
systems, Phys. Rev. Lett. \textbf{121}, 026808 (2018).

\bibitem{Lee} C. H. Lee and R. Thomale, Anatomy of skin modes and topology in non-Hermitian systems, Phys. Rev. B {\bf 99}, 201103(R) (2019).

\bibitem{Yokomizo}K. Yokomizo and S. Murakami, Non-Bloch
band theory of non-Hermitian system, Phys. Rev. Lett. {\bf 123}, 066404 (2019).

\bibitem{Bender} C. M. Bender and S. Boettcher, Real spectra in
non-hermitian hamiltonians having PT symmetry. Phys. Rev. Lett. \textbf{80},
5243 (1998).

\bibitem{SM} See the supplementary materias for more mathematical details of the LE and spectrum sturctures with various parameters.

\bibitem{KZhang} K. Zhang, Z. Yang, and C. Fang, Correspondence between winding numbers and skin modes in non-
hermitian systems, arXiv preprint arXiv:1910.01131 (2019).

\bibitem{Okuma}N. Okuma, K. Kawabata, K. Shiozaki, and M. Sato, Topological Origin of Non-Hermitian Skin Effects, Phys. Rev. Lett. {\bf 124}, 086801 (2020).





\end{thebibliography}
\end{document}